\documentclass[aps,prl,showpacs,superscriptaddress,preprint]{revtex4-2}
\usepackage{amsmath,amsfonts,amssymb}
\usepackage{graphicx}
\usepackage{hyperref}
\usepackage{mathptmx}
\usepackage{times}
\usepackage{bm}
\usepackage{color}
\usepackage{float}

\begin{document}
	\title{Competing Lattice and Defect Dynamics Govern Terahertz-Induced Ferroelectricity in Quantum Paraelectric SrTiO$_3$}
		
\author{L. Cheng}
\affiliation{State Key Laboratory of Electronic Thin Films and Integrated Devices, University of Electronic Science and Technology of China, Chengdu 610054, China}
\author{K. Hu}
\affiliation{State Key Laboratory of Electronic Thin Films and Integrated Devices, University of Electronic Science and Technology of China, Chengdu 610054, China}
\affiliation{Institute of Electronic and Information Engineering, University of Electronic Science and Technology of China, Dongguan 523808, China}
\author{S. Yang}
\affiliation{State Key Laboratory of Electronic Thin Films and Integrated Devices, University of Electronic Science and Technology of China, Chengdu 610054, China}
\author{Yan Liang}
\affiliation{Beijing National Laboratory for Condensed Matter Physics, Institute of Physics, Chinese Academy of Science, Beijing 100190, China}
\author{Jiandi Zhang}
\affiliation{Beijing National Laboratory for Condensed Matter Physics, Institute of Physics, Chinese Academy of Science, Beijing 100190, China}
\author{J. Qi}
\email{jbqi@uestc.edu.cn} 
\affiliation{State Key Laboratory of Electronic Thin Films and Integrated Devices, University of Electronic Science and Technology of China, Chengdu 610054, China}

\begin{abstract}
Intense terahertz (THz) pulses induce transient inversion-symmetry breaking in quantum paraelectric SrTiO$_3$, yet the underlying mechanism remains controversial. Using fields up to $\sim$1.1\,MV/cm, we reveal spatially inhomogeneous THz-field-induced second harmonic generation (TFISH) governed by competing lattice and defect dynamics. Short-lived coherent antiferrodistortive (AFD) modes suppress dipole correlations within $\sim$5\,ps, while heavily damped soft/AFD modes and a defect-induced low-frequency mode ($\sim$0.1-0.3\,THz) jointly prevent long-range ferroelectric coherence in oxygen-vacancy-rich regions. Collective modes manifested by oscillatory TFISH components exhibit softening followed by hardening below a critical temperature $T^*\simeq$28\,K, confirming transient ferroelectric order where defects are sparse. These results reconcile conflicting interpretations, establish defect-mediated competition as a central regulator of light-induced ferroelectricity, and open routes to ultrafast control of quantum materials.
\end{abstract}
\maketitle

Quantum paraelectric oxides such as SrTiO$_3$ (STO) and KTaO$_3$ (KTO) possess incipient ferroelectric (FE) soft modes that soften dramatically with decreasing temperature, yet quantum fluctuations stabilize the centrosymmetric phase down to $T$=0~K~\cite{Cowley1964,Scott_RMP1974,Mueller1979,Fujishita2016}. STO follows the Curie-Weiss law with an extrapolated $T_C\simeq35$\,K and the dielectric constant saturates at a value with an order of 10$^4$ at low $T$~\cite{Mueller1979,Fujishita2016}. KTO displays analogous behavior with a much lower $T_C$($\simeq$4\,K)~\cite{Mueller1979,Fujishita2016}. Similarly, with $T$ decreasing towards to zero, frequency of the soft mode nearly saturates~\cite{Vogt1995,Yamada1969} and deviates from the decreasing trend, which means they do not transform into the FE state but quantum paraelectric (QPE) phase. In order to restore the FE phase, permanent ferroelectricity can be induced by $^{18}$O substitution~\cite{Takesada2006,Itoh1999} or strain engineering~\cite{Haeni2004}, but these methods irreversibly alter the lattice.

Intense terahertz (THz) pulses offer a reversible route to drive transient symmetry breaking in quantum paraelectrics STO~\cite{Li2019,Nova2019,Basini2024,Shin_PRL2022} and KTO~\cite{Li2023,Cheng2023,Yang2025}. THz-field-induced second harmonic generation (TFISH) signals reveal broken inversion symmetry, but their microscopic origin remains disputed: transient long-range FE order~\cite{Li2019,Nova2019}, dipolar correlations among polar nanoregions (PNRs)~\cite{Li2023,Cheng2023}, or hot-phonon effects~\cite{Yang2025} have all been proposed. The unavoidable presence of defects (particularly oxygen vacancies and strain) introduces low-energy collective modes and further complicates interpretation~\cite{2025NatPhy}.

Here, we use THz fields up to $\sim$1.1\,MV/cm to excite STO single crystal at positions with markedly different oxygen-vacancy concentrations. Time- and spatial-resolved TFISH uncovers a dynamical competition: coherent antiferrodistortive (AFD) modes transiently suppress dipolar correlations, while defect pinning and mode damping prevent FE coherence in vacancy-rich regions. Only in clean regions do collective modes exhibit softening-to-hardening transitions below a critical temperature $T^*\simeq28$\,K, revealing light-induced long-range FE order within the quantum paraelectric regime.

\begin{figure*}
	\centering
	\includegraphics[width=\textwidth]{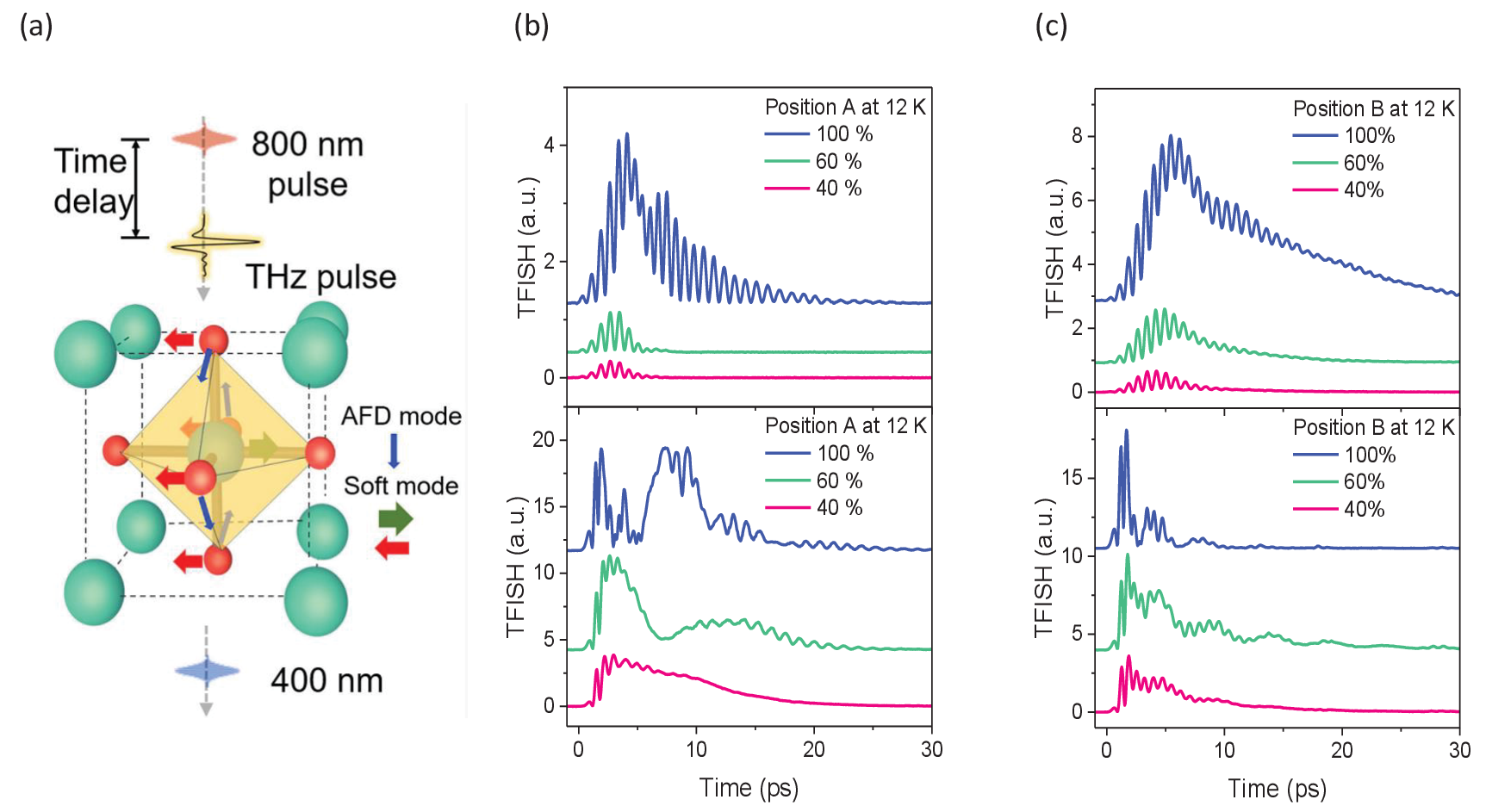}
	\caption{(a) Schematic of time-resolved TFISH. Atomic displacements of soft and AFD modes are also sketched. (b-c) TFISH traces at positions A (low oxygen-vacancy density) and B (high oxygen-vacancy density) at 12\,K under multi-cycle resonant (center frequency of $\sim$0.65\,THz and bandwidth of 0.25~THz) and single-cycle excitation for different field strengths (100\%, 60\%, 40\% of peak fields $E_m^p\simeq220$\,kV/cm and $E_s^p\simeq1.1$\,MV/cm). Striking position dependence highlights the pivotal role of defects.}
	\label{fig:multisingle}
\end{figure*}   

Fig.~\ref{fig:multisingle}(a) illustrates the experimental configuration: a THz pump impulsively drives the lattice, and a delayed 800-nm probe detects transient SHG at 400\,nm~\cite{Gao2024,Hu2023}. Without THz excitation, no SHG is observed at any temperature in the sample. In the tetragonal phase below 110~K, both the soft mode and AFD mode exist~\cite{Vogt1995,Scott_RMP1974}. The former is infrared active and can be resonantly excited via the THz pulse~\cite{Li2019,Nova2019,Shin_PRL2022,Li2023,Cheng2023,Yang2025}. Here, both single-cycle and multi-cycle THz pulses are chosen. Since we only focus on dynamics beyond the THz pulse duration, the time-resolved SHG signals only arise from the second-order susceptibility $\chi^{(2)}(t)$. Given the pivotal role of defects (oxygen vacancies) in this type of study~\cite{Cheng2023,2025NatPhy}, measurements are performed on STO sample with intentionally varied defect densities, prepared to elucidate their impact on the THz-induced phase transition.   

Figs.~\ref{fig:multisingle}(b,c) show representative TFISH traces at 12\,K. At sufficiently strong THz electric fields, both oscillatory and non-oscillatory components are prominent, with dramatic differences between low-defect (position A) and high-defect (position B) regions characterized by Raman spectroscopy~\cite{Chapron2019} (see Supplemental Material). The oscillatory component originates from coherently THz-driven collective excitations and the nonlinear coupling among different modes~\cite{Li2019,Basini2024,Cheng2023,Li2023,Yang2025}. The non-oscillatory component reflects quasi-static broken inversion symmetry due to various debated candidates~\cite{Li2019,Nova2019,Shin_PRL2022,Cheng2023,Li2023,Yang2025,Li2024}. 

The TFISH response exhibits strong nonlinearity with THz field strength. Under multi-cycle excitation, oscillations persist up to $\sim$30\,ps at peak field. With single-cycle pulses, signals at intermediate fields (i.e. $40\%$ of $E_s^p\simeq1.1$\,MV/cm) are similar at both positions, showing decaying non-oscillatory backgrounds superimposed with complex oscillations. At high fields approaching to $E_s^p$, however, behavior diverges dramatically: within the first $\sim$5\,ps the non-oscillatory component is strongly suppressed at both positions, but re-emerges prominently thereafter only at position A, remaining nearly absent at position B. This reveals competing, time- and defect-dependent mechanisms governing the stabilization of transient inversion symmetry breaking.

\begin{figure*}
	\centering
	\includegraphics[width=\textwidth]{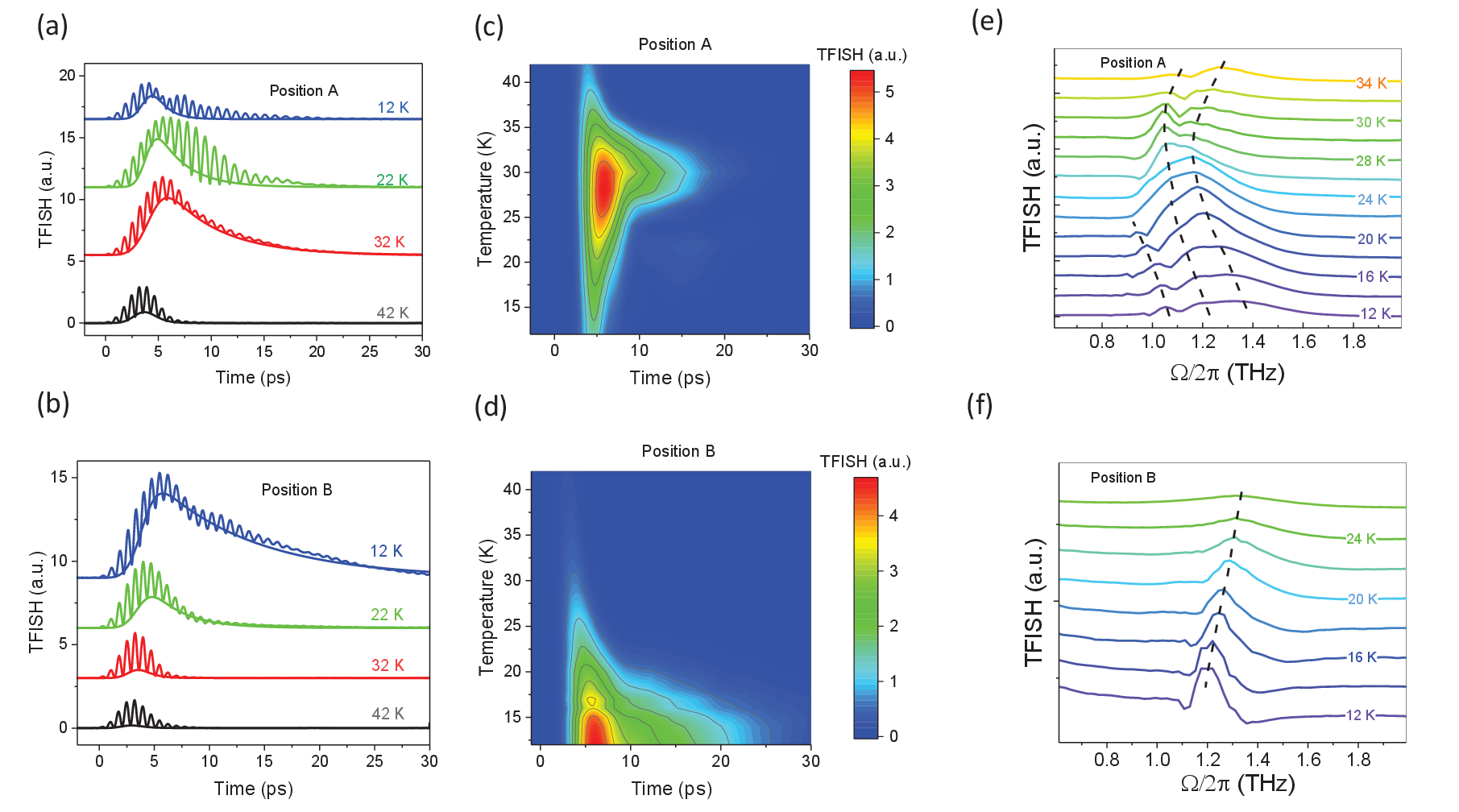}
	\caption{Temperature evolution of TFISH under resonant multi-cycle excitation ($E_m^p \simeq 220$\,kV/cm) with a pulse duration of $\tau^m_{\text{THz}}$($\sim$5~ps). (a),(b) Time-domain traces at positions A and B for several typical temperatures. The solid lines indicate the non-oscillatory components approximated using the exponential decay fitting~\cite{Cheng2023}. (c),(d) The non-oscillatory component after $\tau^m_{\text{THz}}$, attributed to the second-order susceptibility $\chi^{(2)}(t)$, as a function of temperature in the time domain. (e),(f) Fourier spectra of the oscillatory components extracted after $\tau^m_{\text{THz}}$. Position A exhibits clear maxima near $T^*\simeq 28$\,K for the non-oscillatory component, and subsequent softening-to-hardening of collective modes, while position B shows monotonic behaviour characteristic of defect-dominated dynamics. Dashed lines are guide for eyes.}
	\label{fig:multicycle}
\end{figure*}

Multi-cycle resonant excitation provides a cleaner probe of temperature-dependent dynamics (Fig.~\ref{fig:multicycle}). At position A, the non-oscillatory TFISH amplitude and lifetime rise rapidly with decreasing temperature, peak at $T^*\simeq 28$\,K, and then slowly decrease (Figs.~\ref{fig:multicycle}(a,c)) -- behaviour incompatible with continuous increasing as $T$ reduces via mechanism such as PNR correlations or hot-phonon heating alone~\cite{Cheng2023,Yang2025}. Therefore, additional competing effect should appear below $T^*$. Fourier spectra of the oscillatory component (Fig.~\ref{fig:multicycle}(e)) reveal multiple coherent modes. Below $\sim$35\,K, dominant weight appears near twice the soft-mode frequency ($2\nu_s \sim 1.3$\,THz), which softens on cooling and abruptly hardens below $T^*$. This hardening mirrors the growth of a FE order parameter observed in $^{18}$O-substituted STO~\cite{Takesada2006,Itoh1999} and constitutes direct evidence of a transient FE phase emerging in the pristine QPE regime.  

By contrast, position B -- rich in oxygen vacancies -- shows no such anomaly (Figs.~\ref{fig:multicycle}(d,f)). Amplitude and lifetime of the non-oscillatory component grow monotonically with decreasing temperature, and all modes soften continuously without reversal. Specifically, all the collective modes are strongly damped at low temperatures in this region, as demonstrated by the time-domain oscillations after the THz pulse duration in Fig.~\ref{fig:multicycle}(b). Such damped situation dramatically lowers the efficiency of coherent THz driving of both soft and AFD modes, preventing both the nonlinear displacement needed for ferroelectricity~\cite{Li2019,Nova2019,Shin_PRL2022} and the transient lattice distortion that suppresses dipolar correlations~\cite{Zhong1995,Kornev2006} in vacancy-sparse regions.

The oscillatory TFISH signals in position A scale quadratically with THz electric field as expected from hot-phonon scenario~\cite{Yang2025}, which, however, fails to explain the behavior of non-oscillatory component — diagnostic of broken inversion symmetry (Supplemental Material). Furthermore, under off-resonant multi-cycle excitation (center frequency 0.4 THz), the non-oscillatory TFISH signal roughly scales with the field strength but is nearly insensitive to the resonant condition, yet the characteristic FE anomaly at $T^*\simeq28$~K completely disappears (Supplemental Material). These observation unambiguously demonstrate that the hot-phonon effect should play minor role in the non-oscillatory TFISH signal, and thus the transient symmetry breaking.

Note that Raman characterization averages oxygen-vacancy density over our $\sim$80~$\mu$m probe area (a limited spatial resolution of $\sim$60~$\mu$m), so microscopic inhomogeneity persists even within nominally uniform positions A and B. Consequently, the probed region can exhibit behavior akin to spatial phase coexistence: domains dominated by short-range dipolar PNR correlations coexist with domains supporting THz-induced long-range FE order. Both contributions add to the second-order susceptibility, $\chi^{(2)} = \chi^{(2)}_{\text{FE}} + \chi^{(2)}_{\text{PNRs}}$, and, however, compete in the TFISH signal. In sparsely defective position A, their comparable magnitudes yield the observed anomaly -- non-oscillatory amplitude and lifetime increase with cooling initially attributed to dipolar PNR correlations, peak at $T^*\simeq28$\,K as FE order emerges, then decrease due to their competition. In oxygen-vacancy-rich position B, dipolar PNR correlations overwhelmingly dominate, producing monotonic temperature dependence and no detectable FE transition.  

\begin{figure*}
	\centering
	\includegraphics[width=\textwidth]{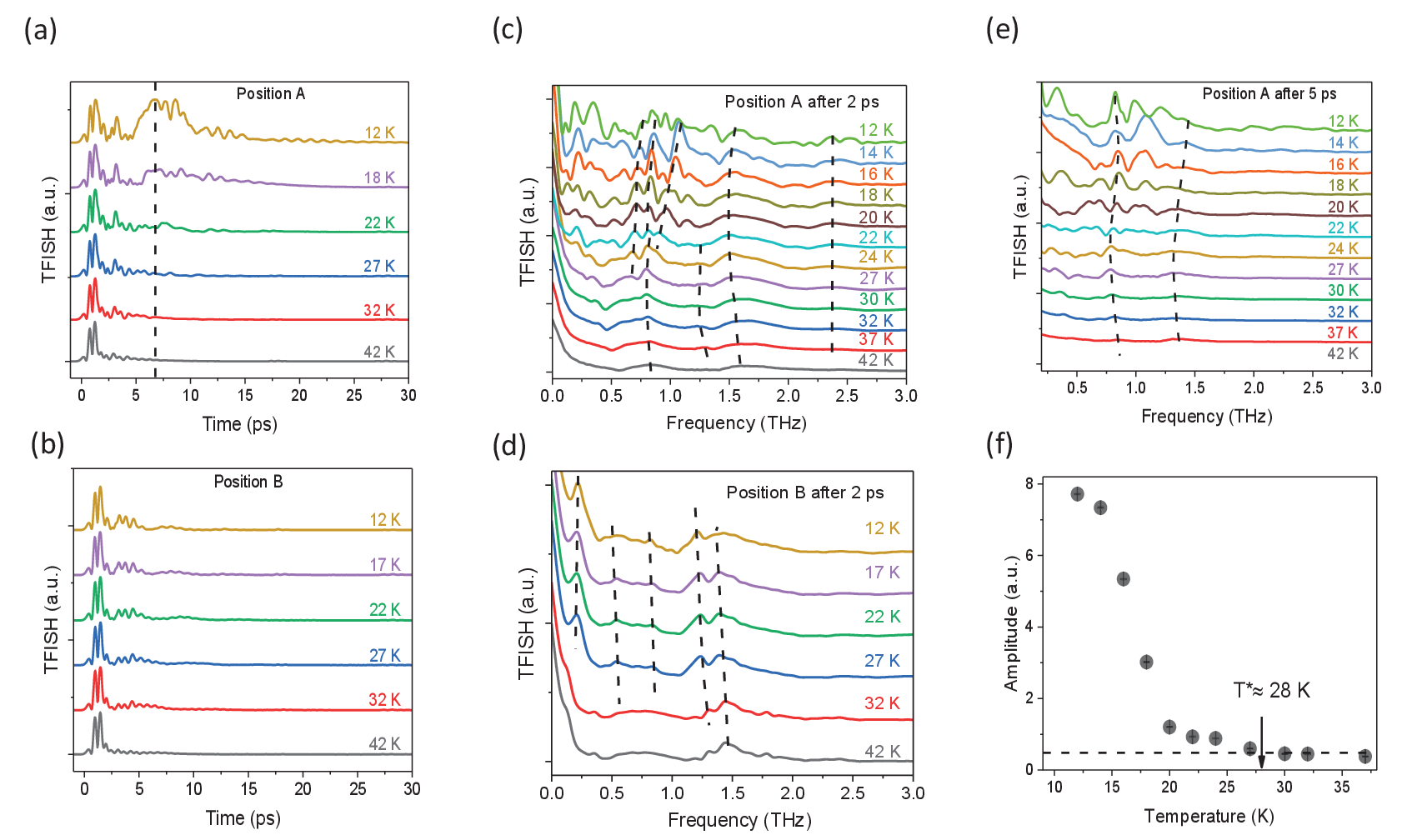}
	\caption{ Single-cycle excitation at $E_s^p \sim 1.1$\,MV/cm with a THz pulse duration of $\tau^s_{\text{THz}}$($\sim$2~ps). (a-b) Time-domain TFISH traces at several typical temperatures in positions A and B. (c),(e) Fourier spectra after $\tau^s_{\text{THz}}$ (left) and after 5\,ps (right) in position A, respectively. (d) Fourier spectra after $\tau^s_{\text{THz}}$ in position B. (f) The amplitude of non-oscillatory component at time indicated by the dashed line in (a) as a function of temperature. Dashed lines are guide for eyes.}
	\label{fig:singlecycle}
\end{figure*}

Single-cycle pulses at intense fields, e.g. $\sim$1.1\,MV/cm, dramatically alter the TFISH signals in the time domain and indicate potential dynamical competing effects. In specific, within the first $\sim$5\,ps, non-oscillatory component is almost completely quenched at both positions (Figs.~\ref{fig:singlecycle}(a,b)). The Fourier spectra exhibit abundant frequencies with spectral weight extending up to $\sim$2.7\,THz (Figs.~\ref{fig:singlecycle}(c,d)), which is not observed in case of the multi-cycle excitation. Previous investigation reveals large spectral weight of AFD mode and its coupled modes with soft phonon in STO appearing between 0.6~THz and 2.6~THz via the nonlinear phonon coupling~\cite{Li2019}. Since the Ti-O-Ti bond-angle distortions via AFD modes cause weakening of dipole–dipole interactions and lead to transiently suppress both FE condensation and dipolar PNR correlations~\cite{Zhong1995,Kornev2006} (the initial quench of non-oscillatory component can be intuitively understood). Considering both soft mode ($Q_s$) and AFD mode ($Q_a$), we can use the coupled oscillators to simulate phenomenologically such suppression in the time domain, evidenced by $Q_s(t)$ developing gradually from an unstable state approaching to a steady nonzero value and the effective potential of the soft mode evolving between single-well and double-well types due to influence of $Q_a(t)$ ~\cite{Mueller1979,Bednorz1984,2024NatMa,Gu2018} (see details in Supplemental Material). 

After $\sim$5\,ps the AFD coherence decays (Figs.~\ref{fig:singlecycle}(a,e,f)), releasing the transient suppression of dipolar correlations and allowing symmetry-breaking signals to recover. In defect-sparse regions (position A), a delayed non-oscillatory component re-emerges and exhibits a distinct two-step temperature dependence below $\sim$28\,K ($T^*$): a slow rise followed by a much steeper escalation below $\sim$20~K (Fig.~\ref{fig:singlecycle}(f)), accompanied by collective modes showing anomalies near $T^*$ (Figs.~\ref{fig:singlecycle}(c,e))--confirming the onset of transient FE order. 

However, The $T$-dependent amplitude reflects subtle competition between the nascent FE phase and persisting dipolar PNR correlations. The FE contribution remains relatively weaker than that under resonant multi-cycle pumping because it is re-established only after the decay of AFD coherence, which has a lifetime larger than the THz pulse duration ($\tau_{THz}^s$). Consequently, short-range dipolar PNR correlations, which strengthen rapidly on cooling together with the potential hot-phonon effect, grow faster than the nascent FE order. This prevents the sharp peak and subsequent decline observed under multi-cycle excitation (Fig.~\ref{fig:multicycle}(c)) and leads to PNR dominance at low temperatures.  

At position B, however, the delayed recovery is essentially absent. This is consistent with observation that the THz-triggered coherent oscillations are strongly damped and survive only within $\sim$5~ps (Fig.~\ref{fig:singlecycle}(b)). Their frequencies undergo monotonic changes as a function of temperature (Fig.~\ref{fig:singlecycle}(d)), suggesting no phase transition. Notably, a low-frequency mode ($\sim$0.1-0.3\,THz) emerges at low temperatures and can persist for $\sim$30~ps, as evidenced by the distinct slow oscillations in time-domain (Fig.~\ref{fig:multisingle}(c)). This low-energy mode hardens either with THz field increasing (Fig.~\ref{fig:multisingle}(c)) or upon cooling (Fig.~\ref{fig:singlecycle}(d)). Its characteristics--including frequency value, temperature-dependent stiffening and confinement to oxygen-vacancy-rich regions--identify it as a defect-induced collective mode associated with PNRs~\cite{2025NatPhy}. Our findings suggest that strongly and coherently excitation of this mode will prevent the establishment of global FE coherence and inhibit the dipolar PNR correlations. 

\begin{figure*}
	\centering
	\includegraphics[width=0.7\textwidth]{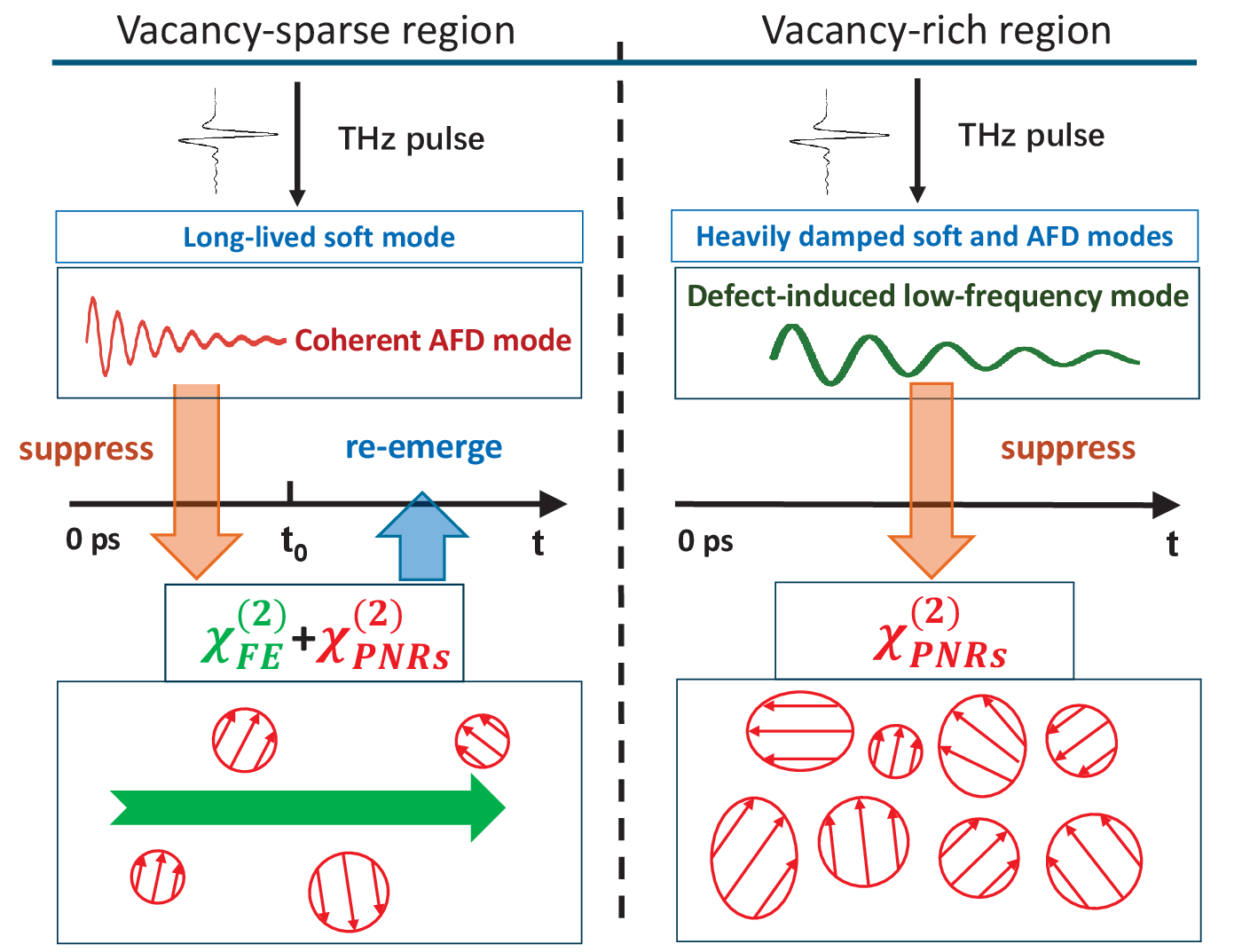}
	\caption{Competing THz-driven pathways in quantum paraelectric SrTiO$_3$. \textbf{Left (oxygen-vacancy-sparse region, position A)}: Exciting coherent AFD rotations transiently suppress dipolar correlations within $t_0$ (orange arrow), e.g. $t_0\simeq$5~ps for single-cycle pulse with $E_s^p$=1.1~MV/cm. After decay, transient ferroelectricity driven by coherent soft-mode and THz-induced correlations of PNRs re-emerge (blue arrow). Non-oscillatory TFISH signals include competing contributions determined by the second-order susceptibility tensors: $\chi^{(2)}_{\text{FE}}$ and $\chi^{(2)}_{\text{PNRs}}$. \textbf{Left (oxygen-vacancy-rich region, position B)}: Soft and AFD modes are heavily damped; a persistent defect-induced low-frequency mode ($\sim$0.1-0.3~THz, dark-green oscillation) dominates and frustrates global coherence, preventing transient ferroelectricity and suppressing dipolar correlations of local polar clusters (PNRs). Non-oscillatory TFISH signals are dominantly contributed by $\chi^{(2)}_{\text{PNRs}}$. }
	\label{fig:summary}
\end{figure*}

Although intense broadband single-cycle pulses inevitably generate a large non-equilibrium phonon population in both positions, the hot-phonon scenario~\cite{Yang2025} should play minor role: it predicts the TFISH signal undergoing a quadratic increase with the THz electric-field, which in stark contrast to the universal early-time quenching and spatial-dependent recovery in the TFISH data at high fields. This is consistent with the results of multi-cycle excitation. 

Our experiments reveal several competing THz-driven processes in STO (schematic in Fig.~\ref{fig:summary}): (i) Short-lived coherent AFD modes ($\lesssim5$\,ps) universally suppress broken inversion symmetry immediately after intense single-cycle THz excitation. (ii) In oxygen-vacancy-rich regions, heavily damped soft/AFD modes reduce driving efficiency, while a persistent defect-pinned mode ($\sim$0.1-0.3\,THz) frustrates both long-range FE order and dipolar correlations of PNRs. (iii) In oxygen-vacancy-sparse regions, collective modes exhibit softening-to-hardening below $T^*\simeq$28\,K, confirming the THz-induced ferroelectricity, which coexists and competes with the dipolar correlations of PNRs. These results reconcile divergent literature interpretations. The THz-induced transient FE phase~\cite{Li2019,Nova2019,Shin_PRL2022} emerges only when defect density is low and AFD suppression decays. PNR-dominated~\cite{Li2023,Cheng2023} or hot-phonon~\cite{Yang2025} scenario can describe defective or lower-field regimes but fail under clean, high-field conditions.

In summary, intense THz excitation initiates a dynamical hierarchy in which lattice distortions, mode damping, and defect pinning compete on picosecond timescales. By mapping this competition at the single-crystal level, we demonstrate that defects are active regulators--not passive scatterers--of light-induced quantum criticality, paving the way for reversible, ultrafast engineering of emergent order in perovskite oxides and beyond. 

We would like to thank Mengkun Liu for helpful discussions, and Ge He for re-confirming usage of Raman spectroscopy to characterize the distribution of oxygen vacancies in STO. This work was supported by the National Key R\&D Program of China (Grant Nos. 2022YFA1403000 and 2024YFA1408502), the National Natural Science Foundation of China (Grants Nos. 12434003, 92365102, 62027807, and 12474107), and the Guangdong Basic and Applied Basic Research Foundation (Grant No. 2024A1515011600).     

	\bibliography{sto}
	
\end{document}